\documentclass{mn2e}

\input psfig

\def\simless{\mathbin{\lower 3pt\hbox
   {$\rlap{\raise 5pt\hbox{$\char'074$}}\mathchar"7218$}}}   
\def\simgreat{\mathbin{\lower 3pt\hbox
   {$\rlap{\raise 5pt\hbox{$\char'076$}}\mathchar"7218$}}}   

\title[Detecting hot Jupiters in FU Orionis objects]
{The spectroscopic signature of hot Jupiters \\ in FU Orionis objects}

\author[C.J. Clarke \& P.J. Armitage]
       {Cathie J. Clarke$^1$\thanks{email: {\tt cclarke@ast.cam.ac.uk}} and Philip J. Armitage$^{2,3}$ \\
       	$^1$Institute of Astronomy, Madingley Road, Cambridge CB3 0HA, UK \\
	$^2$JILA, University of Colorado, 440 UCB, Boulder CO 80309-0440, USA \\
	$^3$Department of Astrophysical and Planetary Sciences, University of Colorado, 
	    Boulder CO 80309-0391, USA}

\begin{document}

\maketitle

\begin{abstract}
We show that if FU Orionis objects harbour hot Jupiters embedded
in their discs, the resulting non-axisymmetric dissipation profile
in the disc would be manifest as time-dependent distortions in the
absorption line profiles of these objects. In order to affect the
infrared line profiles, planets must lie within $\sim 0.5$ au of
the central star, whereas only planets within $\sim 0.1$ au would
influence the optical line profiles. The timescale for modulation
of the line profiles is relatively short (months) in each case,
so that the effect could not have been discovered from published 
spectra (which combine data taken in different
observing seasons). The detection of hot Jupiters in FU Orionis
objects would be in line with the expectations of tidal migration
theories (which predict a high incidence of close planets around
young stars) and would also lend support to models which link the
triggering of rapid rise FU Orionis events to the existence of
a close massive planet.
\end{abstract}


\section{Introduction}
Radial velocity surveys show that the frequency of massive planetary 
companions to late F and G dwarf stars on the main sequence is at 
least 6~percent. Around 1~percent of surveyed stars possess `hot Jupiters', 
that is, massive planets within a tenth of an au of their parent star 
(Butler et al. 2000; Marcy \& Butler 2000). Planetary migration 
theories (Lin \& Papaloizou 1986; Lin, Bodenheimer \& Richardson 1996) 
suggest that the abundance of such planets could be considerably higher
in pre-main sequence stars, but even the main sequence numbers imply 
that a significant fraction -- though probably not a majority -- of 
pre-main sequence (T Tauri) stars harbour planetary companions at 
relatively small orbital radii. Directly detecting planets 
around T Tauri stars is, however, problematic. The rapid
rotation of young stars means that searches for planetary transits
have to contend with additional variability due to rotational modulation
on a timescale of a few days (Aigrain, private communication), whilst
the strong veiling of stellar absorption features in Classical T Tauri
stars have made it difficult to detect even stellar companions by
radial velocity methods (Mathieu 1994).

In this paper, we point out that the subset of accreting pre-main 
sequence stars known as FU Orionis objects present a novel opportunity 
for searching for hot Jupiters. FU Orionis objects are
believed to be episodes of enhanced accretion on to young stars, during 
which the emission from the central star is completely swamped by
the accretion luminosity (corresponding to accretion rates of
$\sim 10^{-5}-10^{-4} M_\odot {\rm yr}^{-1}$; see Hartmann, Kenyon
\& Hartigan 1993; Kenyon 1995; Hartmann \& Kenyon 1996 for
reviews). During the outburst, both the broad band spectral energy distribution 
and the broad double peaked absorption lines of FU Orionis objects are well 
fit by the predictions of a steady state Keplerian accretion disc (Hartmann \& Kenyon 
1985, 1987; Kenyon, Hartmann \& Hewett 1988; Popham et al. 1996). In Section 2, we 
show that the presence of a planet in the inner disc of an FU Orionis 
object would generate a non-axisymmetric pattern of disc emission, 
due to a high rate of accretion onto the planet from the disc. In consequence, 
the resulting line profiles are likely to be asymmetric and time variable, with 
the extra absorption moving between the red and blue wing of the line as the 
planet orbits. We present simple models of the predicted line profiles in this 
case. In Section 3 we discuss which lines provide the best diagnostics of hot 
Jupiters, and discuss the implications of a detection for planetary migration 
theory and the origin of FU Orionis outbursts.

\section{Non-axisymmetric dissipation and spectral line predictions}
The interaction of a planet embedded within a viscous accretion disc 
has several facets. First, if the planet is sufficiently massive, 
gravitational torques from the planet will clear a gap (an annular 
region in which the surface density is greatly reduced as compared 
to the disc) in the disc (Goldreich \& Tremaine 1980; Papaloizou \& 
Lin 1984). Parameterizing the viscosity $\nu$ using the Shakura \& Sunyaev (1973) 
$\alpha$-prescription, $\nu = \alpha c_s^2 / \Omega$, where $c_s$ is the 
sound speed and $\Omega$ the angular velocity, gap opening occurs 
for a mass ratio $q = M_{\rm p} / M_*$ given by (Takeuchi, Miyama \& Lin 1996),
\begin{equation}
 q_{\rm crit} \simgreat \left( { h \over r} \right)^2 \alpha^{1/2},
\end{equation} 
where $h/r$ is the relative scale height of the disc. In the case 
of FU Orionis outbursts, detailed time-dependent modeling of the 
disc suggests that  
$h/r \sim 0.2$ (Bell \& Lin 1994), while numerical simulations of 
angular momentum transport in discs typically find 
$\alpha \sim 10^{-2}$  (e.g. Stone et al. 1996). The stellar mass 
is probably comparable to a Solar mass or somewhat smaller -- for FU Orionis 
itself Hartmann \& Kenyon (1987) infer $M_* \approx 0.6 \ M_\odot$.
Using these parameters, the critical value for $q$ is around $4 \times 10^{-3}$.
This estimate suggests that because of the increased disc thickness during FU Orionis 
events, Jupiter mass planets around Solar mass stars will no longer be able to open a clean gap,  
and will instead be embedded within disc gas (Clarke \& Syer 1996). 

\begin{figure}
 \psfig{figure=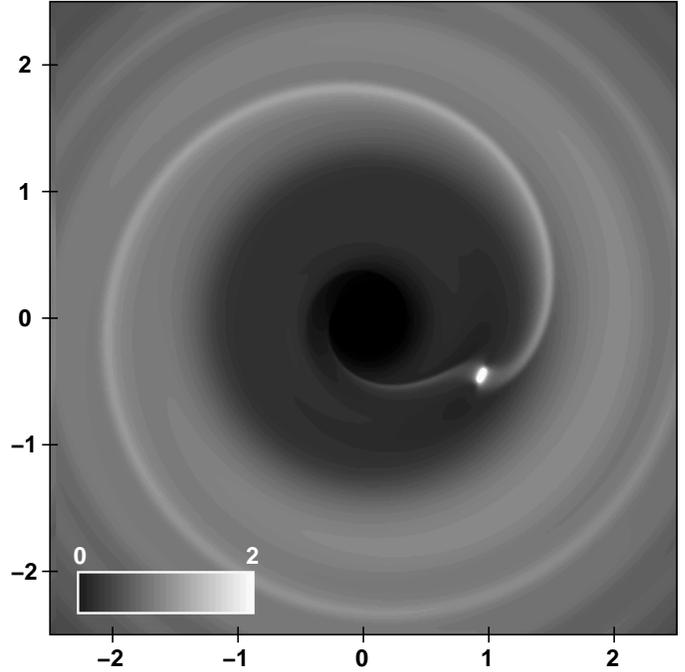,width=3.5truein,height=3.5truein}
 \caption{Surface density from an isothermal, two dimensional hydrodynamic 
 	simulation of the planet--disc interaction, with parameters 
	appropriate for an FU Orionis event. We modeled a planet 
	of mass $3 \ M_J$, on a circular orbit in a disc with $c_s / v_\phi = 0.15$ 
	at the orbital radius of the planet. Angular momentum transport was 
	approximated using a constant kinematic viscosity corresponding to 
	an $\alpha$ of $4 \times 10^{-3}$ at the radius of the planet. 
	The surface density is plotted on a linear scale, with $\Sigma = 1$ 
	(in arbitrary units) corresponding to the initial value of the 
	disc surface density at large radius. To emphasize the structure 
	in the disc, high values of $\Sigma$ close to the planet are 
	suppressed in the image. Scales on the $x$ and $y$ axes are in 
	units of the planetary orbital radius. As discussed in the text, 
	we consider orbital radii $r \sim 0.1 \ {\rm au}$, though this 
	physical scale is not an input for the simulation.}
\end{figure} 

Second, irrespective of the presence or absence of a gap, there can 
be ongoing accretion from the disc onto the planet (Artymowicz \& Lubow 1996). 
The planetary accretion rate can be expressed as a fraction of the 
disc accretion rate,
\begin{equation} 
 \dot{M}_{\rm p} = \epsilon (q) \dot{M}_{\rm disc},
\end{equation}
where $\dot{M}_{\rm disc}$ is the accretion rate in the disc 
at a radius substantially greater than that of the planet. Simulations 
by Lubow, Seibert \& Artymowicz (1999), D'Angelo, Henning \& Kley (2002), 
and Bate et al. (2002) show that the efficiency factor $\epsilon \approx 1$ for 
a planet that is marginally able to open a gap. For larger masses, there is 
an approximately exponential suppression of the accretion efficiency 
(Lubow, Seibert \& Artymowicz 1999). For lower mass planets $\epsilon$ 
also decreases, with a scaling variously measured as $M_p^{1/3}$ from 2D 
simulations (D'Angelo, Henning \& Kley 2002) and as $M_p$ in 3D calculations 
(Bate et al. 2002). In either event, Jupiter mass planets, embedded within 
the inner disc of FU Orionis objects, are expected to accrete mass at a 
significant fraction of the (very high) disc accretion rate.

Finally, the interaction of massive planets with the disc may induce 
eccentricity in both the disc material and the planetary orbit. Whether 
this occurs for Jupiter mass planets is currently uncertain (Papaloizou, 
Nelson \& Masset 2001; Goldreich \& Sari 2003; Ogilvie \& Lubow 2003), 
and is probably dependent upon details of the angular momentum transport 
process in protoplanetary discs (e.g. Ogilvie 2001). We will ignore this 
possibility henceforth.

\begin{table}
\begin{tabular}{lc}
Planet mass & $3 \ M_J$ \\
Stellar mass & $1 \ M_\odot$ \\
Implied disc thickness (at planet radius) & $(h/r) \simeq 0.2$ \\
Effective $\alpha$ (at planet radius) & $\alpha = 4 \times 10^{-3}$ \\
\end{tabular}
\caption{Main parameters of the system shown in Figure~1. Note that 
the simulation can be rescaled to represent other systems with the 
same mass ratio and $h/r$. Likewise, the planetary orbital radius and disc 
accretion rate used are arbitrary.}
\end{table}

To verify some of these estimates, we have simulated the interaction between 
the planet and the disc using the ZEUS hydrodynamics code (Stone \& 
Norman 1992). The most important parameters for the disc and 
planet are summarized in Table 1. 
The simulation was run at a numerical resolution of $240^2$ grid 
points in cylindrical co-ordinates, with outflow boundary 
conditions at the radial edges of the grid. After 120 orbits 
of the embedded planet, by which time 
initial transients have died away, the surface density is as shown 
in Figure~1. As expected based on the arguments given above, the planet 
(here of mass $3 \ M_J$, where $M_J$ is the mass of Jupiter) is 
unable to clear a well-defined gap in a 
disc with relative scale height $h/r$ appropriate to FU Orionis systems, though 
there is a clear drop in the surface density at the radius where a gap 
would form in a cooler disc. A stream from the outer disc carries 
approximately half of the outer disc accretion rate, forming an 
(unresolved) circumplanetary disc and extending toward the inner 
boundary of the simulation.

Apart from gas in the (small) circumplanetary disc, the velocity 
perturbations to the disc flow are negligible. We have therefore 
modeled the influence of the planet upon the line profiles by 
assuming that the dominant effect is due to the enhanced 
dissipation caused by accretion onto the planet. The magnitude 
of the extra dissipation is,
\begin{equation}
 \Delta L = \epsilon { {G M_p \dot{M}_{\rm disc} } \over R_p }
\label{eq_L1} 
\end{equation} 
where $R_p$ is the planetary radius. Since the disc is extremely 
optically thick, this extra luminosity will be radiated over 
a hotspot with a radius $\sim h \approx 0.2 r$. The ratio of the 
effective temperature of the hotspot $\tilde{T}_e$ to the 
effective temperature $T_e$ of the disc is then,
\begin{equation}
 { {\tilde{T}_e^4} \over {T_e^4} } = { {L + \Delta L} \over L },
\label{eq_L2} 
\end{equation} 
where $L$ is the luminosity of the hotspot area in the {\em absence} 
of the enhancement due to the planet. This unperturbed luminosity 
is given by the product of the hotspot area ($2 \pi h^2$, allowing 
for both sides of the disc) and the radiative flux generated 
by a steady-state accretion disc (e.g. Frank, King \& Raine 1992). 
This product yields,
\begin{equation}
 L = { {3 h^2 G M_* \dot{M}_{\rm disc} } \over {4 r^3} }.
\label{eq_L3}
\end{equation}  
From equations (\ref{eq_L1}), (\ref{eq_L2}) and (\ref{eq_L3}), it 
follows that,
\begin{eqnarray}
 { {\tilde{T}_e^4} \over {T_e^4} } \simeq 1 + 
 \left( { \epsilon \over {0.25} } \right) 
 \left( { { M_p / M_* } \over {3 \times 10^{-3} } } \right) 
 \left( { R_p \over {5 R_J} } \right)^{-1} \nonumber \\
 \times \left( { { h / r} \over 0.2 } \right)^{-2}
 \left( { r \over {0.1 \ {\rm au} } } \right),
\label{eq_L4} 
\end{eqnarray}
where $R_J$ is the radius of Jupiter. As noted earlier, a 
disc aspect ratio of $h/r \approx 0.2$ is consistent with 
detailed modeling of FU Orionis events as thermal disc 
instabilities (Bell \& Lin 1994). The planetary radius is 
more uncertain. In isolation, giant planets shrink 
rapidly -- a $3 M_J$ planet contracts below $2 R_J$ within 
a few Myr (Burrows et al. 1997). A planet which is still 
actively accreting will be larger, so for this estimate 
we take a value of $5 R_J$.

For the parameters given in equation (\ref{eq_L4}), we estimate that 
the dissipation rate per unit area is about a factor of two greater 
in the vicinity of the planet than at other phases. For optically thick
emission, this corresponds to a 20~percent increase in the effective
temperature of the disc in the region around the planet. 
In the case of absorption lines whose equivalent width varies only
weakly with temperature (see below) this extra dissipation produces
additional line absorption in the velocity channels centred on the
instantaneous line of sight velocity of the planet. We note that 
the resulting local enhancement of the disc temperature does not 
depend upon the accretion rate through the disc, and (for 
fixed planet radius) becomes more significant at larger 
disc radii.

\begin{figure}
\psfig{figure=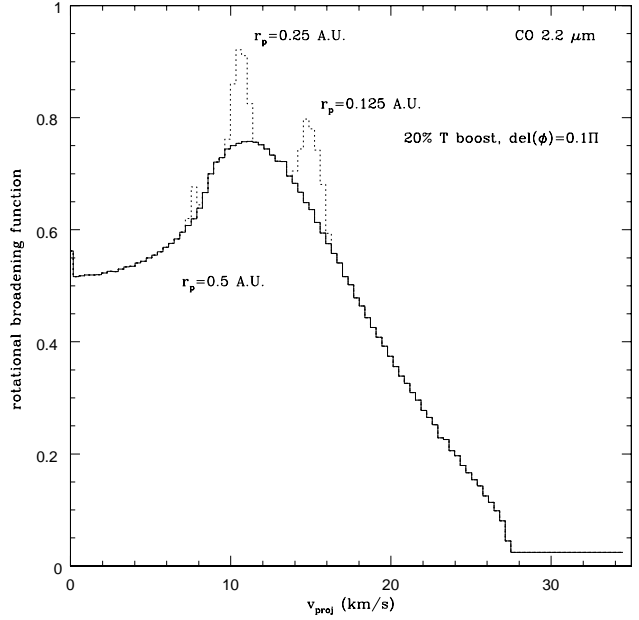,width=3.5truein,height=3.5truein}
\caption{Rotational broadening function for absorption lines in the spectral
region centred on 2.2 microns (see text for details). The normalisation
is arbitrary and half the line is plotted (the other half of the line
would be the mirror image of the unperturbed profile, shown by the
solid histogram). The dashed histograms correspond to models in which
a `planet' is placed at the radii indicated.}
\end{figure}

\begin{figure}
\psfig{figure=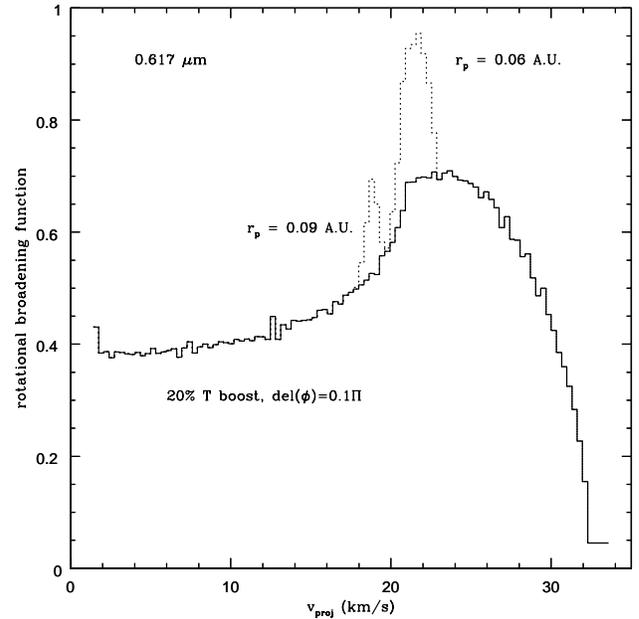,width=3.5truein,height=3.5truein}
\caption{As for Figure~2, except that we show the rotational broadening function 
for the spectral region centred on 6170 \AA.}
\end{figure}

In Figures 2 and 3 we illustrate the distortion of the line profiles
resulting from such a non-axisymmetric surface brightness distribution.
In each case, we have plotted the instantaneous line profiles when the
planet's velocity is most closely aligned to the observer's line of sight.
Half the line is plotted, the other half of the line being the mirror image 
of the unperturbed profile shown by the solid histogram. As the planet
orbits the central star, the feature in the line profile would migrate
back and forth between the red and blue wing of the line 
on the planet's orbital period.  

The quantity plotted is the rotational broadening function (with
arbitrary normalisation) i.e. it is the function with which
the intrinsic spectrum would be convolved in the relevant region of the
spectrum. In practice, the rotational broadening
function in observed spectra of FU Orionis objects is extracted using
cross-correlation techniques, which have the twin advantage of providing
a weighted average over many lines and at the same time being mainly
sensitive to line {\it shape}, rather than line depths
(see Hartmann et al. 1986). In the construction
of Figures~2 and 3 we adopt a simplified model in which the absorption per unit
area at each point in the disc is simply proportional to the
black body emission flux at that point. (i.e. we
neglect variations in the line equivalent width with either
temperature or gravity over temperature ranges appropriate to
the lines in question). A more sophisticated treatment would
involve the summation of lines from appropriate stellar models,
although it is unclear whether this is in fact more correct, given
that in discs there is likely to be significant deposition of mechanical
energy in the atmosphere, which makes comparison with stellar spectra
problematical. In any case, Kenyon et al. (1988) noted that in the infrared,
the saturated nature of the CO lines makes the equivalent widths rather
insensitive to the effective temperature and gravity and a similar
argument can be made for the principal lines in the 6170 \AA \, window
(see discussion in Hartmann and Kenyon 1987).

The two spectral regions illustrated in Figures 2 and 3
correspond to bands centred on
$2.2 \mu m$ and 6170 A, and, following
Hartmann and Kenyon 1987 and Kenyon et al 1988, we have considered
contributions to the line profiles in each spectral region from
the temperature ranges $900-5300$~K and $3300-6600$~K respectively. 
For the temperature
profile of the unperturbed disc, we have adopted the model of Kenyon
et al. (1988) which provided the best fit to the spectral energy distribution
and line profiles of  V1057 Cygni: the disc temperature follows
the classic $r^{-3/4}$ scaling with radius, with maximum temperature of
$\sim 9000$K at the  disc's inner edge ($0.016$ au). Such a disc
temperature distribution corresponds to an accretion rate of
$ \sim 10^{-4} M_\odot {\rm yr}^{-1}$ onto a solar mass star.
Note that material at the inner edge of the disc is too hot to
contribute to the absorption in either of the wavebands shown here
and so there is likewise no contribution from any (presumably hotter)
boundary layer.

In the absence of the planet, the absorption lines exhibit the double
peaked structure familiar from the emission line profiles of cataclysmic
variables. We have here scaled
the velocities so as to roughly reproduce the line widths in
V1057 Cyg, requiring the projected Keplerian velocity at
the stellar surface to be $\sim 40$ km s$^{-1}$ (Kenyon et al. 
1988). The maximum velocity corresponds to the projected
velocity at the inner edge of the line producing region,
whereas the velocity at the line peak reflects the projected
velocity at the location where the line emissivity starts to fall
(close to the Wien cut-off of the relevant transition).
At lower velocities, the line
is produced by a mixture of high velocity material projected at
low velocity and by low velocity material absorbing in the
Wien region of the spectrum. For the parameters employed
here, the  regions of the unperturbed disc that
contribute to the line absorption lie
between $0.025-0.07$ au (for the optical
lines) and $0.04-0.4$ au (for the infrared lines).

We model the presence of the planet as a 20~percent boost in the effective
temperature distributed in a region of azimuthal extent $0.1 \pi$ and
radial half-width $0.1 r_p$ centred on the planet at radius $r_p$, these
choices being motivated by the preceding discussion. 
The effect of the planet is thus manifest in the absorption
line profiles as additional absorption at the projected Keplerian velocity 
of the planet. It is evident from Figures 2 and 3 that the planet produces 
a distortion to the line profile whenever its associated  region of enhanced 
dissipation overlaps the region of the disc producing the relevant line. 
The feature becomes less conspicuous, however, at lower velocities (i.e. 
within the peak of the line profile), since the contribution from the 
region disturbed by the planet is then diluted by the contribution from
higher velocity material projected into this spectral region.

\begin{figure}
\psfig{figure=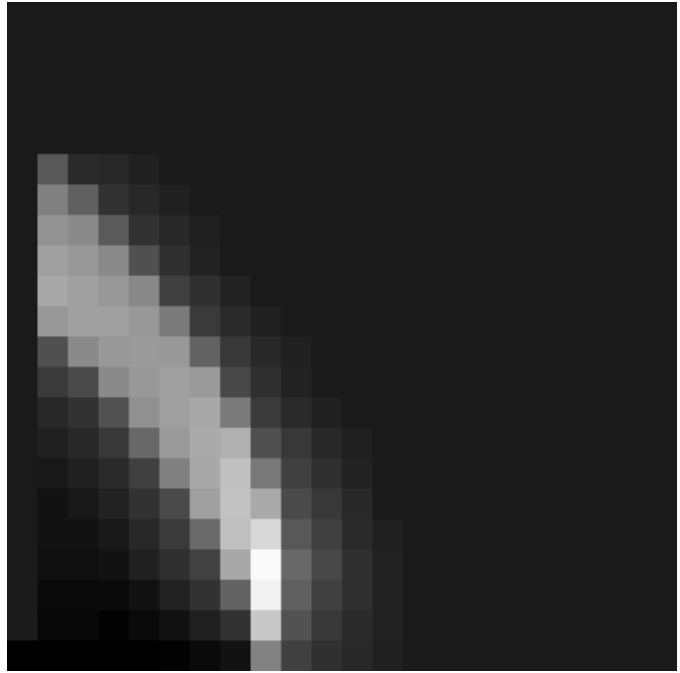,width=3.5truein,height=3.5truein}
\caption{Grey scale representation of the evolution of the rotational broadening
function for the spectral region centred on $2.2 \mu$m. The horizontal
axis is velocity over the range $-8$ km/s (left hand margin) to
$+46$ km/s (right hand margin). Each horizontal slice
is constructed by subtracting the mean (orbit-averaged) value of the
rotational broadening function from its instantaneous value in
that velocity channel. (The black denotes zero difference, whilst the
whitest pixel corresponds to a residual equal to about 10~percent of the
orbit averaged value of the function in that velocity channel). The spectrum
has been degraded to a resolution of $4$ km/s and the vertical (time) axis
corresponds to $3/8$ of the orbital period of the planet.}
\end{figure}

We note that the distortion of the
line profile is quite marked. For example, for the planet located
at 0.25~au, the extra emission from the vicinity of the planet
represents around 22\% of the continuum
at $2.2 \mu$m, produced in a region of the disc from which there is
significant line absorption at this wavelength and in a velocity channel around
$10 $ km/s. Thus the line absorption in this channel is boosted
by this factor. The extra emission from around the planet however
only boosts the overall continuum level at $2.2 \mu $m by 1.5\%
so that the line to continuum ratio in this channel is boosted by
nearly 20\%, as shown in Figure 2. We note that the small contribution
of the planet to the over-all continnum level means that such
a planet would not produce any detectable effect on the broad band
spectral energy distribution of the system.

The features generated by the planet in Figures 2 and 3 are
narrow, with velocity widths -- comparable to the
Keplerian shear across regions of the disc strongly perturbed
by the planet -- of only a few km/s. As such, they would
be hard to resolve in individual spectra, even given the high
spectral resolution of a new generation of infrared spectrometers
such as NIRSPEC and Phoenix. On the other hand, line monitoring should
reveal these features through their regular
motion across the line profile. Figure~4 is a grey scale representation
of the spectral variations produced at $2.2 \mu$m by a planet orbiting
at $0.125$~au, viewed at a spectral resolution of R=75,000. 
Each horizontal slice is a spectrum from which
the mean line profile (over an orbit of the planet) has been subtracted.
The planet's motion is along the line of sight at the bottom
of the plot and the feature moves through the line centre (left
hand margin) as the planet motion passes through the tangent plane.
The characteristic trail produced by the planet is evident
despite the inconspicuousness of the planet's feature in any
individual spectrum.

\section{The detectability of hot Jupiters in FU Orionis systems}

The above crude model suggests that the sort of non-axisymmetric
dissipation profile that one would associate with an embedded
hot Jupiter would have a conspicuous effect on the line profiles
of FU Orionis systems. This immediately raises the question of
whether such a distortion should already have been detected in
FU Orionis objects analysed to date if they indeed harboured such
planets. Two factors however militate against the discovery of
such features in existing spectra: firstly the limited spectral
resolution (particularly in the infrared, where the spectral
resolution of published spectra is comparable with the line
width) and secondly, the fact that in order to enhance the signal
to noise of the lines analysed, spectra have been co-added from
a number of observing runs. We have seen above that the regions
of the disc in which embedded planets would be detectable in this
way are close in (within $0.5$ au) and hence the associated orbital
timescale is short (months). The effect of a feature crossing the
line profile from the red to the blue wing would thus be washed out
by combining spectra from different observing seasons.

In order to detect an embedded planet through its time-dependent
distortion of the line profile, it is of course necessary that the
planet be located in the region of the disc producing the line.
This restricts the technique to the detection of rather close
planets (within $\sim 0.5$ au for an effect to be manifest
in the infrared lines and within $\sim 0.1$ au for a manifestation
in optical line spectra). Such orbital radii are well within
the range of separations of hot Jupiters (to date $37$ and  $18$
exoplanets respectively have been discovered in each of these 
separation ranges\footnote{See e.g. the compilation at {\tt exoplanets.org}, 
though note that many of the planets within 0.1~au have $M_p \sin(i)$ 
less than a Jupiter mass.}). Since the infrared lines are produced in cooler 
regions of the disc, it follows that the infrared lines provide a good 
diagnostic of planets over a larger range of radii.
On the other hand, since the optical lines originate in
a more spatially restricted region of the disc, it is evident
that {\it if} a planet is present at an appropriate radius, it will
be more conspicuous in the optical line profiles, since the distortion
produced by the planet is less diluted by absorption by other annuli.

What, then, is the likelihood of there being an appropriate planet
in a FU Orionis system? If the incidence of planets in
FU Orionis systems follows the statistics of exoplanets around main
sequence stars, then we expect that between 2 and 3 percent of systems 
will harbour a planet within 0.5~au. There would then be little chance 
of there being an exoplanet in the $\approx 16$ systems classified as 
FU Orionis systems to date (Sandell \& Weintraub 2001). There are, 
however, strong theoretical arguments that close planets should
have been commoner during the pre-main sequence stage, and that 
the incidence of such systems on the main sequence represents 
only the (possibly small) fraction that survived migration without 
being swallowed by the star (Lin, Bodenheimer \& Richardson 1996; 
Trilling et al. 1998; Armitage et al. 2002; Trilling, Lunine \& 
Benz 2002). Indeed, given the extreme youth of most FU Orionis 
systems, it is probable that {\em all} close planets present 
in FU Orionis discs are destined to be swallowed, in which case 
main sequence statistics provide no constraint whatsoever on 
the possible population. Although it is impossible to assess by 
what factor the incidence of hot Jupiters is increased in young 
stars, the sign of the effect should motivate the search 
advocated here.

Finally, we remark that FU Orionis objects may be prime targets
for detecting hot Jupiters if, as suggested by Clarke and Syer (1996),
there is a causal link between the two phenomena. Clarke and Syer (1996) 
argued that the most plausible way of triggering a rapid rise outburst 
(i.e. one in which the hundredfold increase in luminosity occurs on a 
timescale of a few months), is if the disc prior to outburst is 
`dammed up' behind such a planet. In the Clarke and Syer model,
the onset of the thermal ionisation instability upstream of the planet
causes the dam to burst, enveloping the planet and causing a rapid rise
in accretion luminosity. It is notable that the planetary characteristics
advocated by Clarke and Syer (a $10$ Jupiter mass planet located at
$\sim 0.1$ au), are just such that they would be detectable through
the absorption line diagnostics discussed here. V1057 Cygni and FU Orionis
(both of which are of the rapid rise variety and which are well studied
spectroscopically) are thus both prime candidates for this sort of
investigation. 

\section*{Acknowledgements}
We thank Doug Lin for useful discussions on the possibility of 
pre-main-sequence hot Jupiters, and 
Mark McCaughrean and Keith Horne for advice 
concerning the observability of the effect. We thank 
the referee for valuable comments that improved the clarity of the 
paper. PJA acknowledges
support from NASA's Origins Program via grant NAG5-13207, and from 
the IoA Visitors Grant. CJC acknowledges
support from the JILA Visitors programme.


\begin{thebibliography}{}

\bibitem{}
 Armitage P.J., Livio M., Lubow S.H., Pringle J.E., 2002, MNRAS, 334, 248

\bibitem{} 
 Artymowicz P., Lubow S.H., 1996, ApJ, 467, L77
 
\bibitem{}
 Bate M.R., Lubow S.H., Ogilvie G.I., Miller K.A., 2003, MNRAS, 341, 213

\bibitem{} 
 Bell K.R., Lin D.N.C., 1994, ApJ, 427, 987
 
\bibitem{}
 Burrows A. et al., 1997, ApJ, 491, 856 

\bibitem{} 
 Butler R.P., Marcy G.W., Vogt S.S., Fischer D.A., 2000, in 
 Planetary Systems in the Universe, eds A.J. Penny, P. Artymowicz, 
 A.-M. Lagrange, and S.S. Russell, IAU Symposium 202, p.~1
 
\bibitem{} 
 Clarke C.J., Syer D., 1996, MNRAS, 278, L23 
 
\bibitem{} 
 D'Angelo G., Henning T., Kley W., 2002, A\&A, 385, 647 

\bibitem{}
 Frank J., King A., Raine D., 1992, Accretion Power in Astrophysics, Cambridge University Press, p.~78
 
\bibitem{} 
 Goldreich P., Sari R., 2003, ApJ, 585, 1024
 
\bibitem{} 
 Goldreich P., Tremaine S., 1980, ApJ, 241, 425 
 
\bibitem{}
 Hartmann L., Hewett R., Stahler S., Mathieu R.D., 1986, ApJ, 309, 275 
 
\bibitem{} 
 Hartmann L., Kenyon S.J., 1985, ApJ, 299, 462
 
\bibitem{} 
 Hartmann L., Kenyon S.J., 1987, ApJ, 312, 243  
 
\bibitem{} 
 Hartmann L., Kenyon S.J., 1996, ARA\&A, 34, 207 
 
\bibitem{} 
 Hartmann L., Kenyon S., Hartigan P., 1993, in Protostars and Planets III, eds 
 E. Levy and J.I. Lunine, University of Arizona Press, p.~497
 
\bibitem{} 
 Kenyon S.J., 1995, Rev. Mex. Astron. Astrophy. Conf. Series, 1, 237 

\bibitem{} 
 Kenyon S.J., Hartmann L., Hewett R., 1988, ApJ, 325, 231  
 
\bibitem{} 
 Lin D.N.C., Bodenheimer P., Richardson D.C., 1996, Nature, 380, 606
 
\bibitem{} 
 Lin D.N.C., Papaloizou J.C.B., 1986, ApJ, 309, 846 
 
\bibitem{} 
 Lubow S.H., Seibert M., Artymowicz P., 1999, ApJ, 526, 1001 
 
\bibitem{} 
 Marcy G.W., Butler R.P., 2000, PASP, 112, 137 
 
\bibitem{} 
 Mathieu R.D., 1994, ARA\&A, 32, 465 
 
\bibitem{} 
 Ogilvie G.I., 2001, MNRAS, 325, 231 
 
\bibitem{}
 Ogilvie G.I., Lubow S.H., 2003, ApJ, 587. 398
 
\bibitem{} 
 Papaloizou J.C.B., Lin D.N.C., 1984, ApJ, 285, 818 
 
\bibitem{} 
 Papaloizou J.C.B., Nelson R.P., Masset F., 2001, A\&A, 366, 263
 
\bibitem{} 
 Popham R., Kenyon S., Hartmann L., Narayan R., 1996, ApJ, 473, 422
 
\bibitem{} 
 Sandell G., Weintraub D.A., 2001, ApJS, 134, 115 
 
\bibitem{} 
 Shakura N.I., Sunyaev R.A., 1973, A\&A, 24, 337
 
\bibitem{}
 Stone J.M., Hawley J.F., Gammie C.F., Balbus S.A., 1996, ApJ, 463, 656 
 
\bibitem{} 
 Stone J.M., Norman M.L., 1992, ApJS, 80, 753 
 
\bibitem{} 
 Takeuchi T., Miyama S.M., Lin D.N.C., 1996, ApJ, 460, 832 
 
\bibitem{} 
 Trilling D.E., Benz W., Guillot T.,
 Lunine J.I., Hubbard W.B., Burrows A., 1998, ApJ, 500, 428
 
\bibitem{} 
 Trilling D.E., Lunine J.I., Benz W., 2002, A\&A, 394, 241

\end{thebibliography}
\end{document}